\documentclass[pra,10pt,twocolumn,groupedaddress,floatfix,showpacs]{revtex4-1}

\usepackage[utf8]{inputenc}  
\usepackage[T1]{fontenc}     
\usepackage[british]{babel}  
\usepackage[sc,osf]{mathpazo}
\usepackage[scaled=0.86]{berasans}  
\usepackage[colorlinks=true, citecolor=blue, urlcolor=blue]{hyperref}  
\usepackage{graphicx} 
\usepackage[babel]{microtype}  
\usepackage{amsmath,amssymb,amsthm,bm,amsfonts,mathrsfs,bbm} 

\usepackage{xspace}  
\usepackage{pgfplots}
\usepackage{xcolor,colortbl}
\usepackage{array}
\usepackage{bigstrut}

\begin{document}



\title{Strong supremacy of quantum systems as communication
resource}


\author{Maria Quadeer}
\affiliation{Optics \& Quantum Information Group, The Institute of Mathematical Sciences, HBNI, C.I.T Campus, Tharamani, Chennai 600 113, India.}

\author{Manik Banik}
\affiliation{Optics \& Quantum Information Group, The Institute of Mathematical Sciences, HBNI, C.I.T Campus, Tharamani, Chennai 600 113, India.}

\author{Andris Ambainis}
\affiliation{Center for Quantum Computer Science, Faculty of Computing, University of Latvia, Raina bulv. 19, Riga, LV-1586, Latvia.} 

\author{Ashutosh Rai}
\affiliation{Center for Quantum Computer Science, Faculty of Computing, University of Latvia, Raina bulv. 19, Riga, LV-1586, Latvia.}


\begin{abstract}
We investigate the task of $d$-level random access codes ($d$-RACs) and consider the possibility of encoding classical strings of $d$-level symbols (dits) into a quantum system of dimension $d'$ strictly less than $d$. We show that the average success probability of recovering one (randomly chosen)  dit from the encoded string can be larger than that obtained in the best classical protocol for the task. Our result is intriguing as we know from Holevo's theorem (and more recently from Frenkel-Weiner's result [\href{http://link.springer.com/article/10.1007/s00220-015-2463-0}{Commun. Math. Phys. {\bf 340}, 563 (2015)}]) that there exist communication scenarios wherein quantum resources prove to be of no advantage over classical resources. A distinguishing feature of our protocol is that it establishes a stronger quantum advantage in contrast to the existing quantum $d$-RACs where $d$-level quantum systems are shown to be advantageous over their classical $d$-level counterparts.
  
\end{abstract}

\pacs{03.65.Ud, 02.50.Le, 03.67.Ac}

\maketitle

\section{Introduction}\label{sec1}
Information theory fundamentally deals with the problem of reproducing at one point, either exactly or approximately, a message selected at another point. The mathematical model for classical information theory was founded in a seminal paper by Claude Shannon in the year $1948$ \cite{Shannon'48}. In the last few decades, application of quantum theory to information processing tasks has lead to several discoveries \cite{Bennett'92,Bennett'93,Grover'96,Shor'97,Ambainis'03} and their synthesis has given birth to the epoch of quantum information theory \cite{Nielsen'00, Wilde'13}. Suitable use of resources from quantum theory can outperform their classical counterparts in several information theoretic, communication, and computational tasks and some of these proposals have already been implemented by the present day quantum technologies \cite{Experiment1,Experiment2}. 

The advantage of quantum mechanical systems over classical systems may be ascribed to the fact that the state of a quantum system is given by a unit vector in some complex vector space. However, some  information processing tasks in quantum theory are restricted. For example, the Holevo's theorem \cite{Holevo'73} puts a restriction on the amount of classical information that can be extracted from a quantum state (\emph{accessible information}). More recently, a remarkable theorem due Frenkel and Weiner \cite{Frenkel'15} shows that the classical information storage in a $d$-Level quantum system can not be more than the corresponding $d$-state classical system. Despite of such restrictions, a quantum system can give advantage in a suitably designed communication task known as \textit{Random Access Codes} (RACs). These were initially introduced by Weisner by the name of conjugate coding \cite{Wiesner'83}, and was later rediscovered in \cite{Ambainis'99,Ambainis'02} by Ambainis \emph{et al.} As pointed out in \cite{Ambainis'02}, the possibility of encoding infinite amount of classical information in a single quantum state (a vector in a complex Hilbert space) and the freedom to perform different non-commutative measurements for extracting the encoded information render quantum random access codes (QRACs) advantageous. QRACs establish that Bennett’s first law of quantum information, i.e., ``$1$ qubit $\succeq 1$ bit" \cite{Schumacher} (here $X\succeq Y$ reads as ``$X$ can do the job of $Y$") is actually strict, i.e., as a communication resource $1$ qubit outperforms $1$ bit in RAC tasks. Recently, Tavakoli \emph{et al.} have studied RACs with high-level symbols that use $d$-level classical and quantum systems \cite{Tavakoli'15}. It has been proved that high-level quantum systems provide significant advantage in the average performances of the RAC tasks over their classical counterparts \cite{Tavakoli'15,Ambainis'15}. In other words, generalized version of Bennett’s first law, that ``$1$ qudit $\succeq 1$ dit" is also strict, i.e., in a $d$-level RAC quantum system outperforms its classical counterpart.

Here, we ask whether a $d'$-level quantum system can outperform a $d$-level classical system in some communication task, where $d'<d$. Interestingly, we find that the answer is in the affirmative. We show that for high-level RACs there exist quantum codes that use relatively lower-level quantum systems for encoding to give better average success probability compared to the best classical codes that use high-level classical systems for encoding. This establishes an advantage of lower level quantum systems as communication resources over a higher level classical system. It is important to note that the quantum supremacy established by our protocol is stronger than that established in the existing QRAC protocols. In other words, we can say that in certain communication tasks $1$ qu-$d'$-dit $\succ 1$ $d$-dit with $d'<d$.

The paper is organized as follows: We briefly review RACs in section-\ref{sec2}; in section-\ref{sec3}, we discuss the high-level RACs; in section-\ref{sec4}, we present our protocol for implementing high-level RACs with lower-level quantum systems, and finally give our conclusions in section-\ref{sec5}.

\section{Random Access Codes: a quick overview}\label{sec2}
Random access codes (RACs) are a class of communication tasks involving two separated parties, (say) Alice and Bob. Alice is given an $n$-bit string $x=x_1...x_n$ chosen uniformly at random from the set $\{0,1\}^n$. The other party, Bob is given a number $y\in\{1,...,n\}$, chosen uniformly at random. Bob's task is to correctly guess the $y^{th}$ bit of Alice. Alice can help Bob in guessing the bit by sending some information about her string. However, the amount of information that Alice can send to Bob is restricted to $1$-cbit. We denote such a RAC by the symbol $[n\xrightarrow{p} 1]$, meaning that $n$ bits are encoded into $1$ bit and $p$ denotes the merit of the success of recovering initial bits which can be either `average success probability' ($P_a$) or `worst case success probability' ($P_w$). In the quantum version of RAC, Alice can encode her $n$-bit string into a two level quantum system, i.e., a qubit. One can also define a more general version of RAC denoted by  $[n\xrightarrow{p} m]$ (with $m\le n$), which is defined analogous to the $[n\xrightarrow{p} 1]$ RACs \cite{Ambainis'99,Ambainis'02,Nayak'99}. Throughout this paper we consider RACs with $m=1$.

For the simplest $2\xrightarrow{p} 1$ RAC, the worst case classical success probability is $P_w^C=1/2$ and the corresponding optimal classical average success probability is $P_a^C=3/4$. Here we use the upper-index $`C'$ to denote the classical case. If Alice sends one of the bit in her string (either first or second bit), Bob can guess that bit correctly but he has to guess the other bit randomly. This naive protocol gives the aforementioned optimal classical success probability (both worst and average), and moreover one can show that no other classical protocol can do better \cite{Ambainis'15}. Interestingly, the authors in \cite{Ambainis'99,Ambainis'02} have shown that for such a classical RAC non-trivial quantum protocols exist which is described as the following : Alice can encode her string $x_1x_2$ into the state $|\psi\rangle\in\mathbb{C}^2$ of a two-level quantum system using the following encoding scheme :
\begin{equation*}
x_1x_2\rightarrow |\psi_{x_1x_2}\rangle:=\sigma_X^{x_1}\sigma_Z^{x_2}|\psi_{00}\rangle,
\end{equation*}
where $\sigma_X,\sigma_Z$ are Pauli operators and $|\psi_{00}\rangle:=(|0\rangle+|0_X\rangle)/\sqrt{2+\sqrt{2}}$, with $|0\rangle$ $(|0_X\rangle)$ being the `up' eigenstate of Pauli-$Z$ (Pauli-$X$) operator. To decode the first bit, Bob, after receiving the encoded particle from Alice, performs a $\sigma_Z$ measurement (i.e performs measurement in computational basis $\{|0\rangle,|1\rangle\}$) on it and guesses the bit value to be $0$  $(1)$ upon obtaining the `up' (`down') outcome. For the second bit, Bob performs $\sigma_X$ measurement and similarily guesses the outcome. For this encoding-decoding scheme, the worst case success probability is $P^Q_w=\frac{1}{2}\left(1+\frac{1}{\sqrt{2}}\right)>\frac{1}{2}=P^C_w$. It is easy to see that the average success probability is the same as that of the worst case, thus $P^Q_a=\frac{1}{2}\left(1+\frac{1}{\sqrt{2}}\right)>P^C_a$. 
Later, Chuang (as mentioned in \cite{Ambainis'02}) generalized this protocol to a $3\xrightarrow{0.78} 1$ QRAC, while Hayashi \emph{et al.} \cite{Hayashi'06-1} proved that no quantum $n\xrightarrow{p} 1$ RAC exists for $n\ge 4$ with $p_w>1/2$. However, it has been shown that $n\xrightarrow{p} 1$ QRAC is possible for any $n\ge 1$ with $p_w > 1/2$ provided that shared randomness is accessible to both parties \cite{Ambainis'09}. 

At this point we would like to mention that QRACs have been originally studied in the context of quantum finite automata \cite{Ambainis'99,Ambainis'02}. They also have applications in quantum communication complexity \cite{Galvao,Klauck,Aaronson}, in particular for network coding \cite{Hayashi'06-2} and locally decodable codes \cite{Kerenidis,Wehner}. QRACs have also been applied to the quantum state learning problem \cite{Aaronson-2}. In recent times, this study also finds applications in semi-device-independent random number expansion \cite{Li'11} as well as semi-device-independent key distribution \cite{Pawlowski'11}. Also, the study of QRACs and it's variant, namely \emph{parity oblivious} RACs has foundational implications \cite{Spekkens'09, Banik'15, Rai'16}, in particular they have been studied for operational depiction of preparation contextuality of mixed quantum states. 

\section{High-level Random Access Codes}\label{sec3}
Recently Tavakoli \emph{et al.} have studied high-level RACs in Ref.\cite{Tavakoli'15}. Alice receives an $n$-dit string $x=x_1,...,x_n$, uniformly at random, where a $dit$ i.e $x_i$ takes values from an alphabet set $\{0,1,...,d-1\}$. She then encodes her string $x$ into a classical $d$-level system (or a quantum $d$-level system), which she can send to Bob in any state she wishes. Bob receives a number $y\in\{1,...,n\}$ chosen uniformly at random and his task is to recover the $y^{th}$ dit, i.e., $x_y$ of Alice's string. 
Such a task is denoted by $[(n,d)\rightarrow 1]$ RAC. As conjectured in \cite{Tavakoli'15} and later proved in \cite{Ambainis'15}, the maximum average success probability for classical RACs is achieved by the `majority-encoding identity-decoding' protocol. A closed analytical formula for the classical average success probability in this task is hard to derive for general values of parameters $n$ and $d$. 
However, for $[(2,d)\rightarrow 1]$ and $[(3,d)\rightarrow 1]$ cases, the analytic expressions can be obtained for the maximum success probabilities, that read $P_{a}^C(2,d)=\frac{1}{2}(1+\frac{1}{d})$ and $P_{a}^C(3,d)=\frac{1}{3}(1+\frac{3}{d}-\frac{1}{d^2})$ respectively. The authors in \cite{Tavakoli'15} have also constructed non-trivial quantum protocols for $[(2,d)\rightarrow 1]$ and $[(3,d)\rightarrow 1]$ cases. The $[(2,d)\rightarrow 1]$ quantum protocol is a generalized version of the $[(2,2)\rightarrow 1]$ protocol already discussed in the previous section. We describe the protocol here for completeness: 
Consider the computational basis $\mathbb{B}_{\mathcal{C}}:=\{|l\rangle\}_{l=0}^{d-1}$, in the Hilbert space $\mathbb{C}^d$ and also consider the Fourier basis $\mathbb{B}_{\mathcal{F}}:=\{|e_l\rangle=\frac{1}{\sqrt{d}}\sum_{k=0}^{d-1}\omega^{kl}|l\rangle\}_{l=0}^{d-1}$, with $\omega$ being the $d^{th}$ root of unity, i.e., $\omega=\exp(\frac{2\pi i}{d})$. Consider the operators $X=\sum_{k=0}^{d-1}|k+1\rangle\langle k|$ and $Z=\sum_{k=0}^{d-1}\omega^k|k\rangle\langle k|$. Alice encodes her strings in the following manner: 
\begin{equation}\label{Hrac}
x_1x_2\rightarrow|\psi_{x_1x_2}\rangle=X^{x_1}Z^{x_2}|\psi_{00}\rangle,
\end{equation}
where $|\psi_{00}\rangle=\frac{1}{N_{2,d}}(|0\rangle+|e_0\rangle)$, with $N_{2,d}=\sqrt{2+2/\sqrt{d}}$ being the normalization constant. For decoding the first dit $x_1$ Bob performs a measurement in the computational basis $\mathbb{B}_{\mathcal{C}}$ and guesses the value as $l$ when the projector $|l\rangle\langle l|$ clicks in the measurement. Similarly, for the second dit, he performs a measurement in the Fourier basis $\mathbb{B}_{\mathcal{F}}$ and guess the dit value according to the measurement outcome. This protocol gives average success probability  $P^Q_a(2,d)=\frac{1}{2}(1+\frac{1}{\sqrt{d}})$ which is strictly greater than the corresponding optimal classical success probability $P^C_a(2,d)=\frac{1}{2}(1+\frac{1}{d})$ for all $d$. As pointed out in \cite{Tavakoli'15}, the optimal advantage of the QRAC over the RAC measured by the ratio of the success probabilities is observed for $d=6$.

\section{High-level RAC with lower-level quantum encoding}\label{sec4}
We consider the $[(2,d)\rightarrow 1]$ RAC task, but Alice has access to a $d'$-dimensional quantum system to encode her classical message where the dimension $d'$ of the quantum system is strictly less than $d$. We investigate whether with a limited dimensional quantum system Alice and Bob can construct non-trivial quantum codes for the task. It turns out that the following QRAC protocol has this interesting feature.
\begin{figure}[t]
\centering
\includegraphics[height=6.5cm,width=8.5cm]{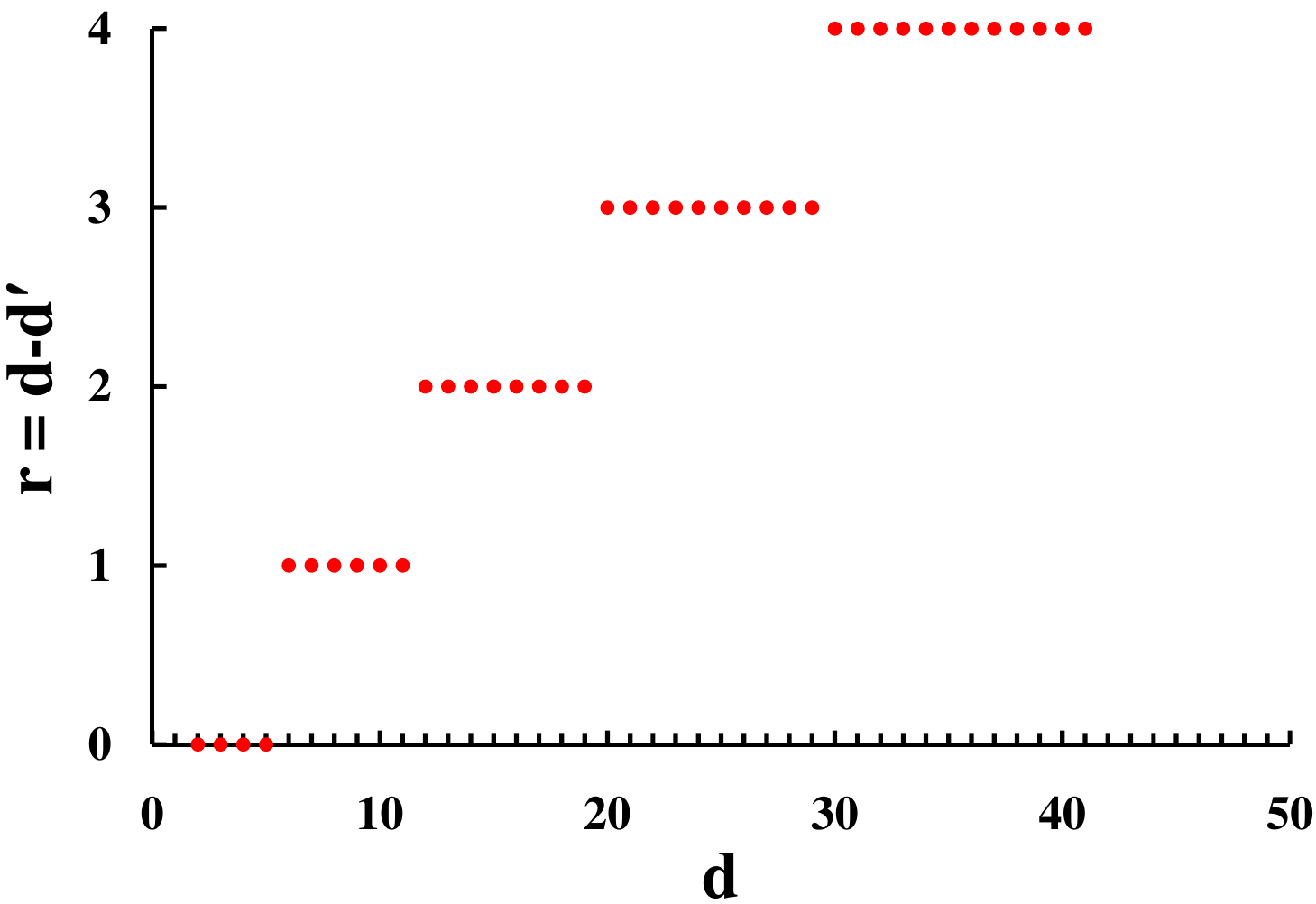}
\caption{(Color online) : Plot $r$ vs $d$. The plot (red dots) gives the \textit{dimensional advantage} $r=d-d'$ of the quantum system used for encoding as a function of $d$. The minimum value of $d$ for which a restricted quantum encoding is advantageous over the best classical $d$-RAC is $d=6$. This advantage in general increases with increasing $d$ although remains flat in some range of $d$. For example, $r=1$ for $d\in\{6,...,11\}$ and then increases to $r=2$ for $d\in\{12,...,19\}$, etc.}
\end{figure}

\emph{Alice's encoding}: Alice encodes her string $x_1x_2\in\{0,1,...,d-1\}^2$ as follows:
\begin{equation}
x_1x_2\rightarrow |\psi_{x_1x_2}\rangle:=\mathcal{G}(x_1,x_2,X,Z)|\psi_{00}\rangle\in\mathbb{C}^{d'}
\end{equation}
with $|\psi_{00}\rangle$ being the normalized state considered in Eq.(\ref{Hrac}), and 
\[
    \mathcal{G}(x_1,x_2,X,Z)= 
\begin{cases}
    X^{x_1}Z^{x_2},& \text{if both}~ x_1,x_2\le d'\\
    \mathbb{1},              & \text{otherwise.}
\end{cases}
\]
with $X,Z$ being the operators as defined in the previous section, and $\mathbb{1}$ is the identity operator.

\emph{Bob's decoding}: To decode the first dit Bob performs measurement in the computational basis $\{|l\rangle\}_{l=0}^{d'-1}$ and on obtaining outcome $l\in \{1,...,d'-1\}$ he guesses the value as $l$, but when he obtains the outcome $l=0$ he guesses an answer from the set $\{0,d',d'+1,...,d-1\}$ uniformly at random. To guess the value of the second dit, he performs a measurement in the Fourier basis $\{|e_l\rangle=\frac{1}{\sqrt{d}}\sum_{k=0}^{d'-1}\omega^{kl}|l\rangle\}_{l=0}^{d'-1}$ and uses the same strategy as before to make a guess. The average success probability of this quantum protocol turns out to be: 
\begin{equation}
P^{Q}_{res}=\frac{d-r}{2d}\left[1+\frac{1}{\sqrt{d-r}}\right],
\end{equation}    
where we call $r=d-d'$ a \textit{dimensional advantage} since it gives the extent to which one can lower the dimension of the quantum system used for encoding and still get advantage over the best classical protocol. For any classical $[(2,d)\rightarrow 1]$ RAC the optimal average success probability is $P^C_a(2,d)=\frac{1}{2}(1+\frac{1}{d})$. Therefore, in order to do better, our quantum protocol must satisfy
\begin{equation}
P^{Q}_{res}>P^C_a(2,d)\qquad\implies d>r^2+3r+1.
\end{equation}
Condition-(4) implies that our quantum protocol gives advantage for $d\geq 6$ and $r\in \left\{ 1,...,\lfloor\frac{1}{2}(-3+\sqrt{4d+5})\rfloor  \right\}$, where $\lfloor a\rfloor$ is the greatest integer less then or equal to $a$. This means that the smallest value of $d'$ that yields an advantage is restricted. For small values of $d$, the allowed value of $d'$ is exactly $d-1$, as dictated by condition-(4) (See Fig.1). As $d$ increases, the allowed values of $d'$ become $d-1$ and $d-2$ and so on. Therefore, one cannot arbitrarily choose any lower $d'$ dimensional quantum system to supersede the optimal classical $d$-RAC protocol. In fact, there exists a minimum value of $d'$, encoding below which would be bad, since it would yield a success probability lower than that of the optimal classical $d$-RAC. Fig.1 illustrates how the dimensional advantage $r=d-d'$ grows with $d$. Note that for $d=\{2,3,4,5\}$ we get no dimensional advantage (In Fig.1, $r=0$ corresponds to this case). 

\section{Concluding Remarks}\label{sec5}
In conclusion, our protocol demonstrates the supremacy of a lower-dimensional quantum system over it's higher-dimensional classical counterpart, a consequence of the existence of superposition of states and non-commutative measurements in quantum theory.
Our result is slightly stronger than the earlier results on RACs which prove the advantage of using a quantum system over a classical system of the same dimension \cite{Tavakoli'15}. In general, the study of QRACs is important as there exist communication scenarios where the use of  quantum resources proves to be of no advantage over their classical counterparts, a result of Holevo's theorem \cite{Holevo'73}, a general version of which has been proved more recently by Frenkel and Weiner \cite{Frenkel'15}. While these theorems address the question of encoding and decoding a random variable, RACs are concerned with decoding randomly chosen parts of such an encoded random variable. Finally, we would like to state that proving the optimality of high-level RACs using a limited dimensional quantum system remains an open problem and needs further investigation. 
\begin{acknowledgements}
MB and AR would like to thank Prof. G. Kar for his comments and suggestions. MB acknowledges his visit at the University of Latvia where a part of the work was done. MQ acknowledges her visit at the Institute of Mathematical Sciences, Chennai. AR and AA acknowledge support by the European Union Seventh Framework Programme (FP7/2007-2013) under the RAQUEL (Grant Agreement No. 323970) project, QALGO (Grant Agreement No. 600700) project, and the ERC Advanced Grant MQC.
\end{acknowledgements} 


\end{document}